# A simple intelligent adaptive network

Mingyang Bai[1] and Daqing Li[1*]

[1] Department of Reliability and Systems Engineering, Beihang University, Beijing 100083, China

* Corresponding author: daqingl@buaa.edu.cn

## Abstract

For real-world complex system constantly enduring perturbation, to achieve survival goal in changing yet unknown environments, the central problem is constantly adapting themself to external environments according to environmental feedback. Such adaptability is considered the nature of general intelligence. Inspired by thermodynamics, we develop a self-adaptive network utilizing only macroscopic information to achieve desired landscape through reconfiguring itself in unknown environments. By continuously estimating environment entropy, our network can adaptively realize desired landscape represented by topological measures. Our network achieves adaptation under several scenarios, including confinement on phase space and geographic constraint. A unique power law distinguishes our network from memoryless systems. Furthermore, our simple strategy could enable brain network and communication network to adaptively maintain essential topological characteristics. Compared to data-driven methods, our self-adaptive network is understandable without careful choice of learning architectures and parameters. Our self-adaptive network could help to understand adaptive intelligence through the lens of thermodynamics.

One of the most mysterious emergent phenomena is intelligence[1-6]. There are some popular definitions regarding the ability of solving difficult problems[2] or cognition process[3] as intelligence. However, these definitions depend on specific scenarios or tasks. Meanwhile, general intelligence is widely believed to be based on adaptability[1,4,5], which is shared by different intelligent systems. To achieve survival goal in changing environments, the intelligent systems must constantly adapt their internal structure to external environments according to environmental feedback with limited information. There are many systems with such adaptive intelligence in nature. By improving vessel conductivity when its flow increases, animals and plants realize highly optimized transport networks adapted to environment[7-9]. After suffering external damage[10] or aging-induced losses[11], brain reorganizes its neural circuits to meet cognitive task demands. For the intelligent adaptive system, the central problem is finding a self-adaptation strategy, which enables system to adjust itself to achieve survival goals in changing environments. For complex system with hierarchical structure, it needs to build the feedback loop bridging the macroscopic system target with the microscopic system elements[12]. Thus, the key requirements are representing macroscopic target and finding a self-adaptation strategy with minimum information, enabling system to adjust its internal multi-scale structure accordingly.

Ashby first modeled adaptive behavior on a mathematical formalism, where the survival goal is defined as maintaining essential variables (such as body temperature or blood oxygen level) within given boundaries[13,14]. It is revealed that the key of adaptation is finding efficient adaptation rules guiding self-organization[13,14]. These rules instruct systems to reconfigure their internal organization when systems do not realize targets. For low-dimension systems, the adaptation process could be achieved through simply random adjustment. As the dimension increases, the efficient adaptation rules become critical. The most famous rules may be Thorndike's law of effect[15] and Hebb's rule[16], which greatly influence reinforcement learning[17] and deep learning[18-20]. Thorndike's law of effect states that behavior leading to desired outcome tends to be repeated more often, which is realized by reconfiguring connection of neurons following Hebb's rule[21,22]. Thus, animals and humans could adjust their behavior in new environments for the desired outcomes. In addition to Hebb's rule for neuronal systems, numerous adaptation rules[23] are proposed for epidemic spreading[24] , transport networks[7-9], social networks[25-28] and physiological networks[29,30]. According to these adaptation rules, the system internal structures co-evolve with system states, inducing many novel dynamic phenomena. However, these adaptation rules are primarily hardcoded for specified targets and systems, lacking general self-adaptation strategies that can be programed with various complex systems. Although data-driven methods such as reinforcement learning demonstrate great power to enable complex systems to adapt to uncertain environments[31,32], they usually require deliberated learning architectures and massive training samples.

To address these challenges, here we develop a simple self-adaption strategy for complex systems to realize desired landscapes, by selectively accepting the transition of system state based only on macroscopic information, where correct acceptance probability $A(\sigma_i \to \sigma_j)$ is learned from history. Inspired by thermodynamics, the goal of our strategy is set to landscape of macrostate, namely the negative logarithm of probability distribution for macroscopic variables of interest, which has been used for depicting material[33], protein[34], cancer[35], ecosystem[36,37] and engineering systems[38]. The valleys of the landscape correspond to the desired target for macrostate. To achieve this target, in our strategy, system transition from one state to another due to environmental disturbance, such as adding/removing an edge in network, will be selectively accepted, where the acceptance probability $A(\sigma_i \to \sigma_j)$ only depends on macrostates before and after transition. We show that suitable acceptance probability can be determined by target landscape and environment entropy (Fig. 1a). In unknown environments, self-adaptive systems need to automatically estimate environment entropy for correct acceptance probability. By extending the Wang-Landau method, we derive a simple method to estimate entropy, enabling systems to continuously adjust acceptance probability according to environmental feedback. It is shown that system will converge to target landscape under our adaptation rule (Fig. 1b). The adaptation process is described by a monotonically decreasing relative entropy corresponding to the Boltzmann $H$ function[39]. The relative entropy decreases with time following unique power law distinguishing our self-adaptive systems from memoryless systems. As a prototypical example, we design an intelligent adaptive network programmed with target landscapes represented by topological measures, including modularity and others. In a wide variety of unknown environmental conditions, our network shows the capacity to realize desired landscape by reconfiguring itself with only macroscopic information. We demonstrate that our intelligent adaptive network could change from one target landscape to another, a capacity known as transformability, as a crucial mechanism for complex system resilience[36,40]. Furthermore, we test our strategy in real networks, including the brain functional networks and communication networks.

## Theory

**Model.** Consider a complex system of numerous microstates that compose a large phase space $\Gamma = \{\sigma_1, \sigma_2, ...\}$. For example, an undirected network with $n$ nodes and $m$ edges can have $\binom{\frac{n(n-1)}{2}}{m}$ possible microstates. Driven by environment, system continuously transitions from one microstate to another described by a stationary Markov process $T$, where $T(\sigma_i \to \sigma_j)$ represents the transition probability from microstate $\sigma_i$ to microstate $\sigma_j$ at one timestep. We denote the probability of microstate $\sigma_i$ as $p(\sigma_i)$. When accepting all transitions driven by environment, the distribution of system microstate converges to the stationary distribution $p(\sigma_i) = p^{env}(\sigma_i)$, where the superscript 'env' represents that system is dominated by environment.

**Target**. Due to the difficulty of analyzing countless microstates, thermodynamics coarse-grain different microstates with the same macroscopic property into one macrostate. Inspired by thermodynamics, we only focus on quantity of interest $x(\sigma_i)$ for each microstate $\sigma_i$, thus one macrostate $x$ corresponds to a set of microstates $\{x(\sigma_i) = x | \sigma_i \in \Gamma\}$. For adaptive behavior, we consider the fundamental survival goal[13,14] of maintaining essential macroscopic variables $x(\sigma_i)$ within given boundaries. For complex systems, we focus on the distribution of macrostate

$$p(x) = \sum_{x(\sigma_i)=x} p(\sigma_i), \tag{1}$$

which is the sum of probability of microstate with same macrostate $x$. The negative logarithm of stationary distribution -ln[$p(x)$] corresponds to free energy landscape[33], which depicts the macroscopic property of complex systems. Not limited to physical and chemical systems[33,34], macroscopic landscape also describes the ecosystem resilience[36,37] and engineering systems[38]. Moreover, because macroscopic landscape allows fluctuation and one macrostate contains numerous microstates, macroscopic landscape as target is more feasible for complex systems under uncertainty compared to microscopic target[41]. Our goal is to realize desired landscape $U^{design}(x)$, thus the target stationary distribution $p(x)$ is

$$p^{design}(x) \propto \exp\left(-U^{design}(x)\right). \tag{2}$$

Thus the valley of target landscape $U^{design}(x)$ represents the desired region for essential macrostate. Without any intervention, the system landscape will be dominated by environment. Namely, when the system accepts all transition driven by environment (represented by Markov process $T$), the stationary distribution $p(x)$ of macrostate $x$ becomes

$$p^{env}(x) = \sum_{x(\sigma_i)=x} p^{env}(\sigma_i) = \exp\left(-U^{env}(x)\right), \tag{3}$$

where $p^{env}(x)$ is the stationary distribution of macrostate under Markov process $T$. $U^{env}(x) = -\ln\left[p^{env}(x)\right]$ is the environment-dominated landscape, which corresponds to entropy[42]. Below we seek to steer system from environment-dominated landscape $U^{env}(x)$ to target landscape $U^{design}(x)$ by self-adaptation (Fig. 1b).

**Realizing target landscape with adaptation rule**. To realize desired macroscopic landscape, we seek to selectively accept the transition of system state to change stationary distribution. While selective acceptance is used in Markov chain Monte Carlo for realizing desired microstate distribution with microscopic information[43], this information is usually inaccessible for complex systems with numerous microstates. To realize desired macrostate distribution with limited information, under detailed balance condition, we derive the suitable acceptance probability

$$A\left(\sigma_i \to \sigma_j\right) = \min\left(1, \frac{\exp\left(U^{design}(x_i) - U^{design}(x_j)\right)}{\exp\left(U^{env}(x_i) - U^{env}(x_j)\right)}\right) \tag{4}$$

when system is driven from microstate $\sigma_i$ with macroscopic property $x_i = x(\sigma_i)$ to microstate $\sigma_j$ with macroscopic property $x_j = x(\sigma_j)$. Exploiting only macroscopic information $x_i$ and $x_j$, system could reconfigure the stationary distribution to $p(x) = p^{design}(x)$ by selectively accepting environment drive. See derivation in **Methods**.

In unknown changing environments, system use its estimation $\widehat{U}^{env}(x)$ to replace $U^{env}(x)$ in Eq. (4), where $\widehat{U}^{env}(x)$ should be continuously updated based on environmental feedback. However, finding a proper way to update $\widehat{U}^{env}(x)$ is challenged by non-stationary transition probability, which arises from ever-changing acceptance probability when $\widehat{U}^{env}(x)$ updates. Considering correspondence between $U^{env}(x) = -\ln[p^{env}(x)]$ and entropy[42], we extend the Wang-Landau method[44,45] and derive a simple method for updating estimation $\widehat{U}^{env}(x)$ while pursuing target. Our method degrades into the Wang-Landau method when $U^{design}(x)$ is constant. According to our method, at each timestep, the estimation $\widehat{U}^{env}(x)$ on the current macrostate $x_i$ will be updated as

$$\widehat{U}^{env}(x_i) = \widehat{U}^{env}(x_i) - f \times \exp(U^{design}(x_i)), \quad (5)$$

where adaptation rate $f > 0$. The larger $f$, the faster updating while larger fluctuation for $\widehat{U}^{env}(x)$. See derivation in **Methods**. Our adaptation rule, as described in Eqs. (4-5), is understandable. For instance, consider the situation when system visits macrostate $x_H$ too frequently compared to target. According to Eq. (5), $\widehat{U}^{env}(x_H)$ will decrease faster than other $\widehat{U}^{env}(x)$. According to Eq. (4), with low $\widehat{U}^{env}(x_H)$, acceptance probability of escaping macrostate $x_H$ will increase while acceptance probability of entering macrostate $x_H$ will decrease. Subsequently, the frequency of visiting macrostate $x_H$ will decrease. Thus, our adaptation rule establishes a negative feedback loop between the deviation from macroscopic target and acceptance decision for realizing target (see Supplementary Section 1).

Taken together, by implementing our strategy shown in Fig. 2, with only macroscopic information, system could adaptively realize desired landscape in unknown environments. In the following, our strategy is applied to developing an self-adaptive network which realizes different desired landscapes represented by topological measures automatically.

## Results

**Self-adaptive network programed with target landscape.** Consider a complex network under disturbance, environmental disturbance will remove existing edges or create new edges at each timestep (see environment setting in **Methods**). To realize desired macroscopic property represented by target landscape, system needs to continuously adjust its behavior (through acceptance probability) according to environmental feedback. As our first example, we choose modularity $M$, which measures the tendency of network division into densely connected subgroups[46], as macrostate of interest. Our goal is to design a self-adaptive network which starts from environment-dominated landscape $U^{env}(M)$ to desired target landscape $U^{design}(M)$ automatically in unknown environments. To test strategy capability, we choose a bistable landscape as the target landscape (see target setting in **Methods**). Such landscapes with multiple attractors support complex functions in mechanical, biological, and chemical engineering[47].

Employing our strategy, our self-adaptive network gradually adjusts its behavior to realize target landscape in the initially unknown environment (Fig. 3a). As network continuously updates estimation $\widehat{U}^{env}(M)$ (Fig. 3b), the distance to target landscape, quantified by the relative entropy $D_{KL}(q||p^{design})$ between empirical distribution $q(M)$ and target distribution $p^{design}(M)$, decreases monotonically with time following a power law (blue circles, Fig. 3c). Programed with bistable landscape $U^{design}(M)$, self-adaptive network is of bistability. Namely, when $M < M_0$, network modularity will fluctuate around $M_1$ in a long time (Fig. 3d), where edges are comparatively uniformly distributed among nodes; while when system is perturbed to $M > M_0$, network modularity will fluctuate around $M_2$ in a long time, where network is divided into several subgroups (Fig. 3e).

The adaptation process can be described by a universal exponent. As shown in Fig. 3c, two regimes are identified, with a crossover at $\tau_c$, corresponding to the correlation time (see Supplementary Section 3). When $t \gg \tau_c$, $D_{KL}(q||p^{design})$ decreases with time following a power law

$$D_{KL}(q||p^{design}) \propto t^{-\alpha}, \quad (6)$$

where exponent $\alpha \approx 2$. Thus self-adaptive network will gradually approach target landscape. While for non-adaptive systems accepting all environmental disturbance, empirical distribution $q(M)$ will converge to environment-dominated distribution $p^{env}(M)$ instead of target distribution $p^{design}(M)$. Thus relative entropy $D_{KL}(q||p^{env})$ decreases with time for non-adaptive systems (red diamonds, Fig. 3c), following a different power law

$$D_{KL}(q||p^{env}) \propto t^{-1}, \quad (7)$$

which describes how non-adaptive systems converge to environment-dominated landscape $U^{env}(M)$. We derive the exponent in Eq. (7) for purely random systems and stationary Markov systems in Supplementary Section 2. Thus exponent $\alpha > 1$ in Eq. (6) suggests that our adaptive system continuously exploits historical observation to adjust itself to realize target distribution, which can distinguish adaptive systems from memoryless systems. And we find that the exponent $\alpha$ depends on the target landscape. For symmetrical bistable landscape $U^{design}(x)$, the exponent $\alpha \approx 2$ is independent of details. For $U^{design}(x)$ with single minimum, the exponent $1 < \alpha < 2$ decreases as probability becomes concentrated (see Supplementary Section 4).

**Adaptation under confinement on phase space.** Real-world systems suffer many constraints in phase space (red regions in Fig. 4a), which challenges system adaptability. As our second example, we consider such a scenario that some microstates are inaccessible due to environmental restrictions. Specifically, with other setting similar to our first example, we forbid 80% of all possible edges in the second example. These edges could not be added during network evolution. Under confinement, the volume of phase space is reduced to approximately $5^{-200}$ of the original one (see **Methods**). With enormous forbidden regions, self-adaptive network needs to automatically find possible paths for realizing target landscape. Although numerous microstates are forbidden, the macroscopic property represented by environment estimation $\widehat{U}^{env}(M)$ changes slightly, where the minimum of $\widehat{U}^{env}(M)$ is still around $M = 0.45$ (red line, Fig. 4b). With slightly adjusted estimation, our network still realizes target landscape in the new environment (Fig. 4c). The relative entropy $D_{KL}(q||p^{design})$ decreases with time following a power law, where exponent $\alpha \approx 2$ still holds under confinement (red triangles, Fig. 4d). The microscopic evolution of network is demonstrated in Fig. 4e. During evolution, forbidden edges (red lines, Fig. 4e) gradually disappear due to environmental constraints. Employing remaining 20% edges, self-adaptive network still realizes desired bistable target landscape (right panel, Fig. 4e). Such adaptability of our self-adaptive network enables systems to flexibly select suitable microscopic evolution trajectories for realizing target under uncertain confinement.

**Adaptation under geographic constraint**. One important constraint for real-world networks is geographic constraint, where long-range edges require more resources to build and maintain. As our third example, we consider a environment under geographic constraint to test the adaptability of our strategy. Specifically, all nodes of the network are embedded in a two-dimensional square lattice (Fig. 5a). Here, the strength of geographic constraint is represented by a characteristic length $\zeta$. The edge with length $d_{ij} \gg \zeta$ is easy to remove while hard to add (see **Methods**). Since average shortest path length $L$ is significantly affected by geographic constraint, we choose $L$ as macrostate of interest in this example. The target landscape is set to a landscape with one minimum, which is different from previous setting. We employ our self-adaptive network in environments with different geographic constraint strengths $\zeta$. As $\zeta$ decreases, the environment-dominated landscape $\widehat{U}^{env}(L)$ increases slower with $L$ (Fig. 5b), suggesting that environment with small $\zeta$ is more likely to have large $L$. According to Eq. (4), self-adaptive network will reduce the acceptance probability for increasing $L$ with decreasing $\zeta$. Thus network adaptively realizes target under different $\zeta$ (Fig. 5c), which is also described by the power-law decay of $D_{KL}(q||p^{target})$, where exponent $\alpha \approx 1.34$ (Fig. 5d). Under different geographic constraints, self-adaptive network realizes macroscopic target by different microscopic modes. Under weak geographic constraint ($\zeta = 100$), there are many long-range edges (Fig. 5e). As $\zeta$ decreases, the network realizes target landscape with a highly modularized network structure and a few long-range edges (Fig. 5f-g), which has been observed in real-world spatially embedded networks[48].

**Transformability.** Besides adaptively realizing a specific target in various unknown environments, adjusting target landscape to changing requirement is critical for system resilience. Such capacity is defined as transformability[36] (Fig. 6a). As our final example, we investigate the transformability of our self-adaptive network. Here, we choose average clustering coefficient $C$ as macrostate of interest. We sequentially perform a large transformation (Fig. 6b) and a small transformation (Fig. 6e). It is found that our self-adaptive network can successfully transform to new target landscape (Fig. 6c, 6f).

We further investigate historical experience effect on transformation. When target landscape is changed from an old one (Fig. 6b, 6e, dashed lines) to a new one (Fig. 6b, 6e, solids lines), the estimation $\widehat{U}^{env}$ learned in old tasks may help system realize new task faster. To investigate such historical experience effect, we compare the convergence of $D_{KL}(q||p^{target})$ for network with historical experience and network without any historical experience (reset $\widehat{U}^{env} \equiv 0$). For large transformation $U_1 \to U_2$, $\widehat{U}^{env}$ learned in old tasks only helps system in the early state (Fig. 6d). While for small transformation $U_2 \to U_3$, $\widehat{U}^{env}$ learned in old task significantly accelerates target realization in the new task, reducing the time to decrease $D_{KL}(q||p^{target})$ (purple dashed arrow, Fig. 6g). Such difference comes from the distance between old target and new target, which is small in transformation $U_2 \to U_3$ but large in transformation $U_1 \to$ $U_2$. This suggests that environment entropy $\widehat{U}^{env}$ plays the role of world model[2,5,49] enabling system to generalize previous experience to similar tasks. When the current landscape becomes hard to maintain, such transformability of landscape to evolve a new way of living is critical in social-ecological systems[36].

**Application in real network systems**. We apply our self-adaptation strategy to natural or artificial networks for maintaining essential topological characteristics. We first consider the brain functional network, where nodes represent regions of interest and edges represent the functional correlation between regions (Fig. 7a, see network setting in **Methods**). The functional network describes how different brain regions exchange information. To adapt to diverse cognitive demands, the brain network dynamically alternates between the segregation and integration modes[50]. In the segregation mode, the functional network divides into sparsely connected modules, allowing each to process specialized information independently. In the integration mode, the connections between different modules become denser for enhancing information integration across different regions. Our self-adaptation strategy provides a simple way to guide brain network to switch between different modes. We choose modularity as the macroscopic quantity of interest, which describes the strength of segregation. The target is set to a bistable landscape, where the valleys describe the desired modes. Even under random noise (see environment setting in **Methods**), the brain network employed with our strategy could dynamically alternates between integration mode and segregation mode, guided by the macroscopic target landscape (Fig. 7b-d).

We also consider a wireless communication network in which nodes represent sensors, base stations and other equipment, while edges represent the direct communication between nodes. The nodes are randomly deployed within a square area of side length $L$. Communication between nodes is constrained by geographical distance and node degree (see network setting in **Methods**). Wireless communication network needs to be resilient when experiencing environmental disturbance. A good topological measure for local fault tolerance is clustering coefficient[51], which represents the density of triangle in network. For example, if nodes A, B, and C form a triangle, a failure on edge AC can be bypassed using AB and BC as an alternative path (Fig. 7a). This would only increase the communication length by one hop without significantly changing the route. However, maintaining the clustering coefficient around a desired value is particularly challenging for traditional topology control methods[52] due to the heterogeneity of massive elements, including the random spatial distribution of nodes and variations in node communication capacity $r_i$, $k_i^{max}$. We use our strategy to address these challenges. We construct a noise to constantly drive network evolution (see **Methods**). By selectively accepting the endogenous noise, the communication network successfully keeps clustering coefficient fluctuating around a desired value across different geographical strengths (Fig. 7f-h). Our simple self-adaptive strategy endows the networks with adaptive intelligence. By selectively accepting exogenous and endogenous noise, these networks adaptively coordinate vast heterogeneous microscopic elements, ensuring that the macroscopic properties align with the target landscapes.

## Discussion

Adjusting internal organization for survival goals is critical for complex systems in uncertain environments. Such adaptability is believed the essence of general intelligence, which is independent of specific tasks. Previous work on adaptive dynamical networks has primarily focused on adaptation rules based on microscopic information, while often neglecting the guidance from macroscopic survival goal — the necessary information for developing more complex systems, such as life[53]. Here we take a step further by focusing on the adaptation strategy based on thermodynamics. To steer complex systems toward macroscopic goal, we develop a simple intelligent adaptive network. With only macroscopic information, our strategy guides systems to make decisions at microscopic level with removal or addition of edges for achieving desired landscape at macroscopic level. In our strategy, the gap between macroscopic targets and microscopic decision-making is bridged by the entropy of the network macrostate. Most adaptive dynamical networks neglect top-down causation[53-55]. These networks adjust their structure based solely on information at microscopic level, without survival goal at macroscopic level. With only bottom-up causation, some order structures like simple biological molecules, Turing pattern, sand piles and flocks could emerge[53]. While developing more complex systems, such as a single living cell, requires top-down causation[53,55], which contains necessary information about environmental niches. Our strategy provides an example of decomposing macroscopic targets into microscopic decisions through entropy[55].

Our simple intelligent adaptive network has the adaptability to adjust its internal mechanism for its goal. Most prior important work for steering complex systems assumes known dynamics, which is challenged by uncertain environments in the real world. Existing methods usually need to model system dynamics and are hard to apply for complex systems with many degrees of freedom and unknown dynamics. Estimating countless microscopic parameters is difficult for nonlinear complex systems[56,57]. Our strategy represents external uncertain environments as entropy, which is easier to be estimated. By continuously updating estimation of entropy, our intelligent adaptive network could efficiently adapts its behavior to uncertain environments for achieving target landscapes. With the adaptation rule that system adjusts its behavior based on the deviation from macroscopic target, our coarse-grain strategy may open a new route to close the loop across different scales, which is believed key to designing complex systems[58].

Our strategy is physics-inspired. Compared to data-driven methods, our explainable strategy does not require complex architecture and elaborated pre-training. Moreover, our work may help to understand the principle behind complex system adaptive intelligence. We find a universal exponent distinguishing our simple intelligent adaptive network from non-adaptive systems. Our approach suggests that entropy could play the role of world model[2,5,49] for multi-scale systems. Our simple intelligent adaptive network can be applied to designing future intelligent complex technological systems and help understand adaptive intelligence through the lens of thermodynamics.

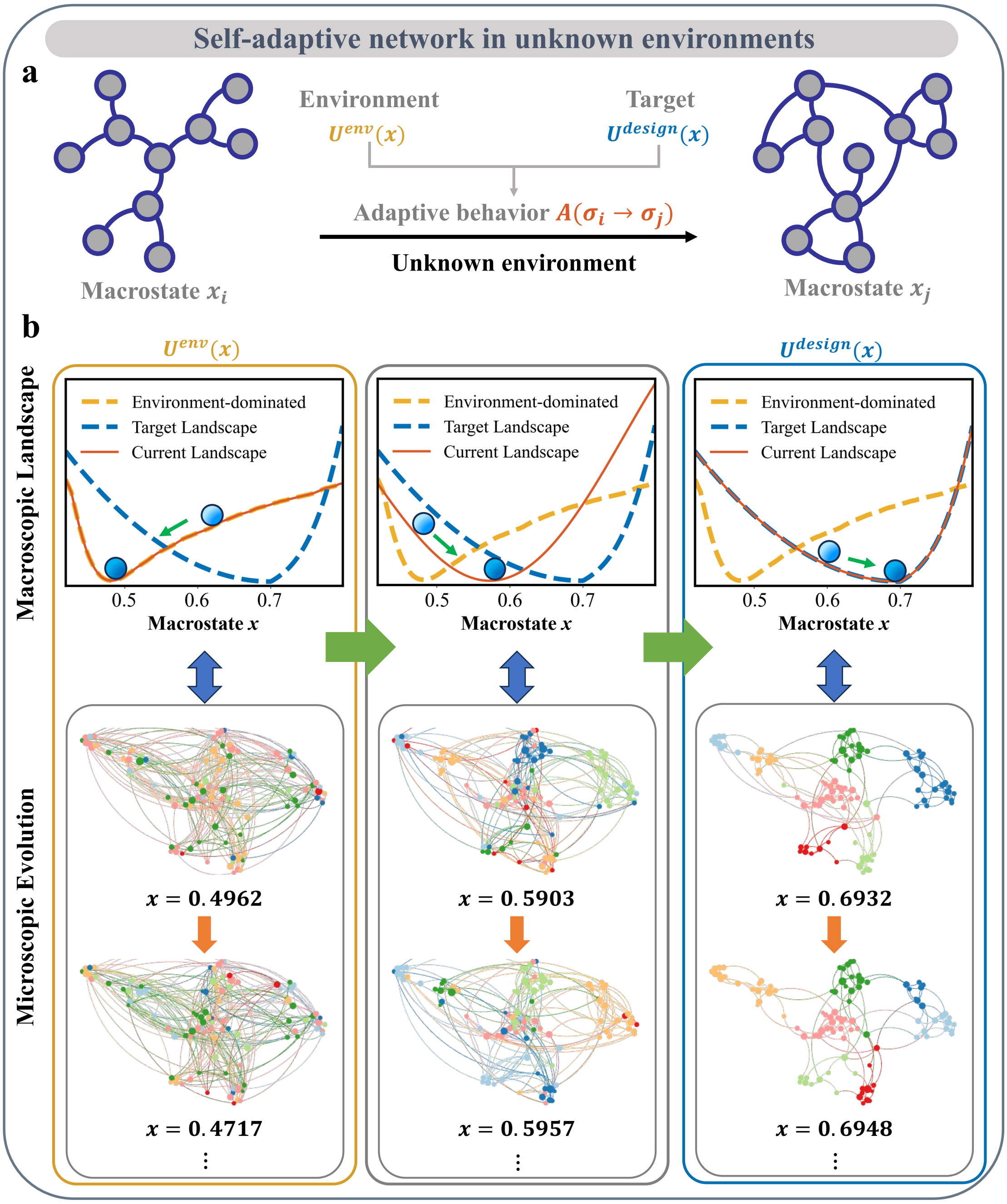

**Fig. 1| Self-adaptive network. a.** Self-adaptive network in unknown environments. Environment will drive network from one microstate to another at each timestep (such as removing or adding edges in network). When accepting all environmental disturbance, network macrostate $x$ is governed by environment-dominated landscape $U^{env}(x)$. To realize target landscape $U^{design}(x)$, our self-adaptive network selectively accept environmental disturbance with suitable acceptance probability $A(\sigma_i \to \sigma_j)$ based only on macroscopic information (macrostates $x_i$ and $x_j$ before and after transition). The suitable behavior $A(\sigma_i \to \sigma_j)$ depends on both environment property $U^{env}(x)$ and target $U^{design}(x)$. In unknown environments, self-adaptive network needs to learn environment property $U^{env}(x)$ and adjust its behavior accordingly. **b.** An example of self-adaptive network. In unknown environments, self-adaptive network adaptively adjusts its behavior (represented by $A(\sigma_i \to \sigma_j)$) based on environmental

feedback. Thus network gradually modifies its landscape from environment-dominated landscape $U^{env}(x)$ to target landscape $U^{design}(x)$. In this example, macrostate of interest $x$ is chosen as modularity. Network gradually forms several communities to increase modularity. Here, nodes in the same community are shown in the same color. Nodes with more connections are larger.

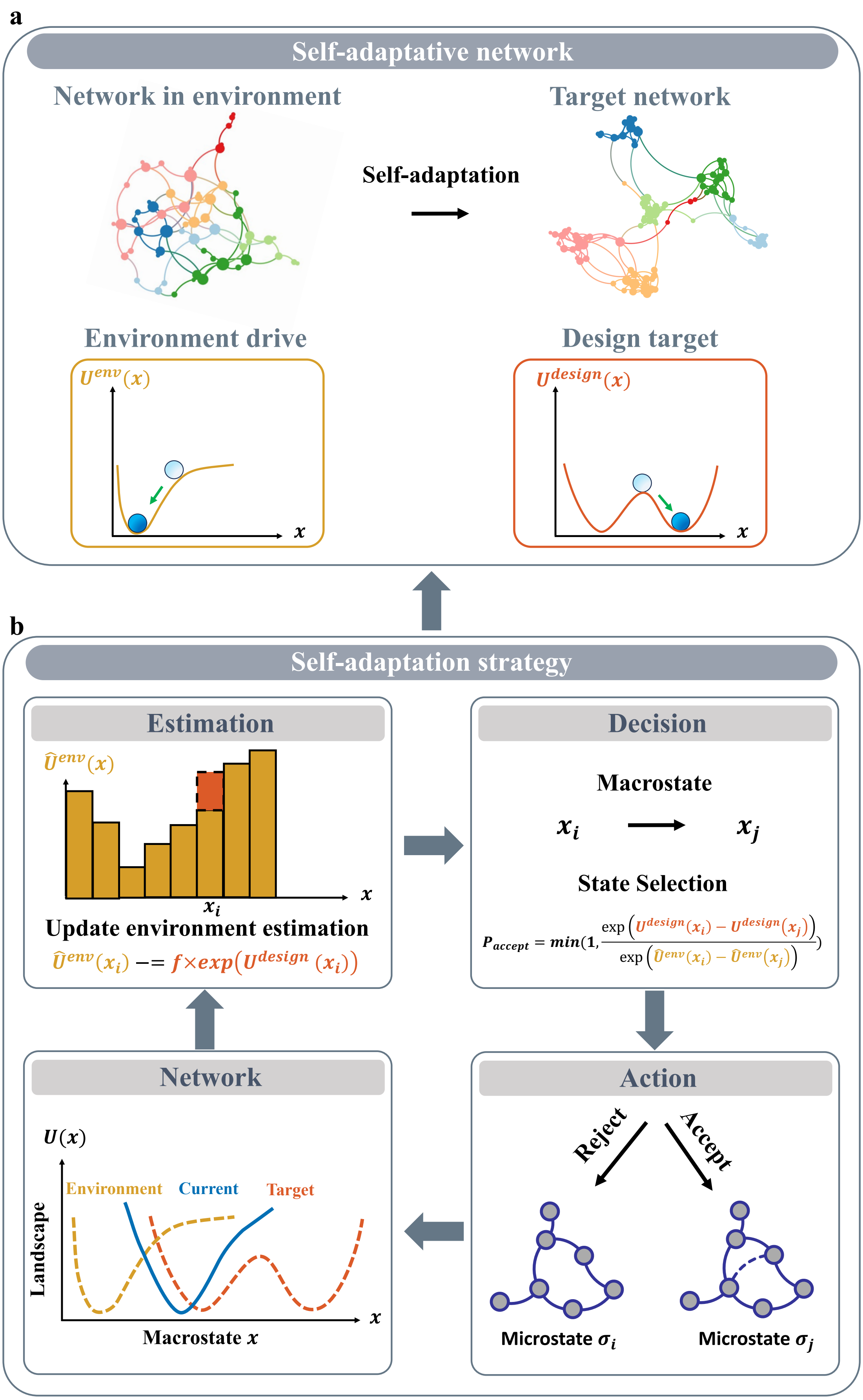
a
Self-adaptative network
Network in environment
Target network
Self-adaptation
Environment drive
Design target
$U^{env}(x)$
$U^{design}(x)$
x
b
Self-adaptation strategy
Estimation
$\hat{U}^{env}(x)$
$x_i$
Update environment estimation
$\hat{U}^{env}(x_i) -= f \times exp(U^{design}(x_i))$
Decision
Macrostate
$x_i$ → $x_j$
State Selection
$P_{accept} = min(1, \frac{exp(U^{design}(x_i) - U^{design}(x_j))}{exp(\hat{U}^{env}(x_i) - \hat{U}^{env}(x_j))})$
Network
$U(x)$
Landscape
Environment
Current
Target
Macrostate x
Action
Reject
Accept
Microstate $\sigma_i$
Microstate $\sigma_j$

**Fig. 2| Self-adaptation strategy for designing self-adaptive network. a.** Self-adaptive network: network structure, unknown environment and target landscape $U^{design}(x)$. Environment will drive system from microstate $\sigma_i$ to microstate $\sigma_j$ with unknown probability $T(\sigma_i \rightarrow \sigma_j)$ at each timestep. When accepting all environment drive, network macrostate $x$ is governed by environment-dominated landscape $U^{env}(x)$ . Self-adaptive network could only observe macrostate $x_i$ for $\sigma_i$ and macrostate $x_j$ for $\sigma_j$ to decide whether accepting the transition. Self-adaptive network reconfigures the network structure to target landscape $U^{design}(x)$ by selectively accepting disturbance in unknown environments. **b.** Self-adaptation strategy. At each timestep, our self-adaptive system will first update its environment estimation, then decide whether accepting environmental disturbance. At estimation stage, system updates environment estimation $\widehat{U}^{env}(x)$ on current macrostate $x_i$ via Eq. (5). At decision stage, to decide whether to accept or to reject transition $\sigma_i \rightarrow \sigma_j$, system calculates the acceptance probability $p_{accept}$ based on environment estimation $\widehat{U}^{env}(x)$ and target landscape $U^{design}(x)$. At action stage, system accepts transition $\sigma_i \rightarrow \sigma_j$ with acceptance probability $p_{accept}$. This action determines the network structure at next timestep, which shapes the macroscopic landscape.

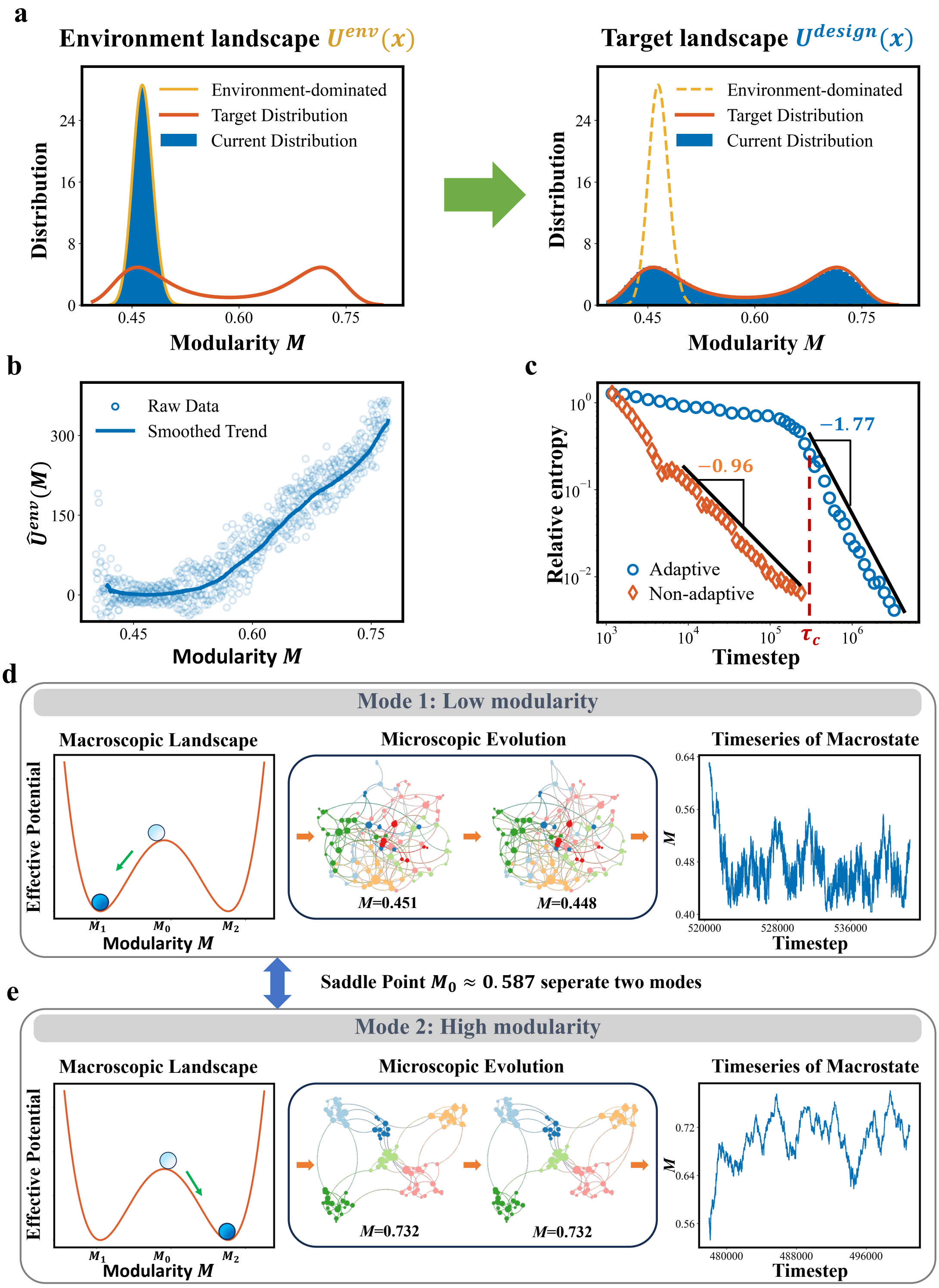

**Fig. 3| Self-adaptive network programed with target landscape. a.** Self-adaptive network realizes target landscape. When accepting all environmental disturbance, the distribution of modularity $M$ is dominated by environment (yellow line, left panel). Employing our self-adaptation strategy, network successfully approaches target distribution (red line, right panel). The empirical distribution $p^{env}(M)$ before and after employing self-adaptation strategy is obtained by simulating $2.42 \times 10^6$ timesteps and $5 \times 10^6$ timesteps, respectively. **b.** Environment estimation $\hat{U}^{env}(M)$ learned from

environmental feedback after $5 \times 10^6$ timesteps. To visualize the trend of $\widehat{U}^{env}(M)$, we smoothed raw data (hollow points) using a moving average method with a window size of 30 to obtain the curve (solid line). The minimum of smoothed $\widehat{U}^{env}(M)$ is adjusted to zero by adding a constant term. **c.** Relative entropy between empirical distribution $q(M)$ and stationary distribution decreases with time following a power law. For our self-adaptive network, stationary distribution is target distribution $p^{design}(M)$. The relative entropy $D_{KL}(q||p^{design})$ decreases with time following a power law (blue circles), where the best-fitting exponent is $\alpha = 1.77 \pm 0.04$, based on the result of one simulation using data points where $D_{KL}(q||p^{design}) < 0.5$. For non-adaptive systems accepting all environmental disturbance, the stationary distribution $p^{env}(M)$ is determined by environment. The relative entropy $D_{KL}(q||p^{env})$ decreases with time following a different power law (red diamonds), where the best-fitting exponent is $\alpha = 0.96 \pm 0.02$, based on the average of 15 independent simulations using data points where $D_{KL}(q||p^{env}) < 0.5$. All the scaling exponents are determined by linear regression in the log-log space. **d-e.** The network macrostate timeseries and microscopic evolution when network switches between low modularity mode (around $M_1 \approx 0.458$) and high modularity mode (around $M_2 \approx 0.715$). Two modes are separated by the saddle points $M_0 \approx 0.587$. See target setting in **Methods**. Here, nodes in the same community are shown in the same colors. Nodes with more connections are larger.

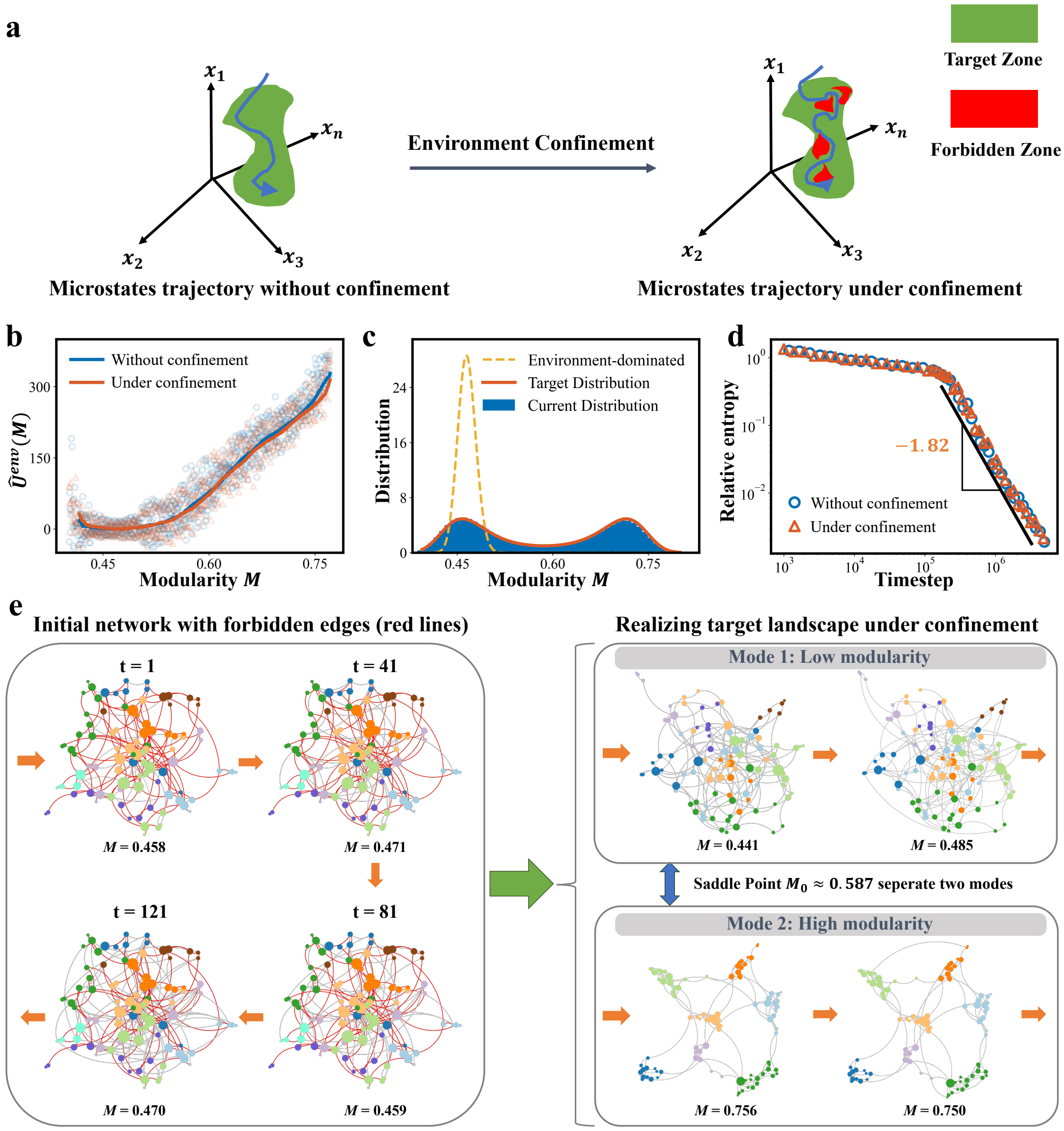

**Fig. 4| Adaptation under confinement on phase space. a.** Network could not access some microstates due to environmental restriction (see environment setting in **Methods**). **b.** Comparison of environment estimation $\widehat{U}^{env}(M)$ under confinement (red triangles and red line) and without confinement (blue circles and blue line). Both estimations are obtained by simulating $5 \times 10^6$ timesteps. To visualize the trend of $\widehat{U}^{env}(M)$, we smoothed raw data (hollow points) using a moving average method with a window size of 30 to obtain the curve (solid lines). The minimum of smoothed $\widehat{U}^{env}(M)$ is adjusted to zero by adding a constant term. **c.** The distribution of network modularity $M$ for self-adaptive network under confinement before and after employing self-adaptation strategy. The empirical distribution before and after employing self-adaptation strategy is obtained by simulating $1 \times 10^6$ timesteps and $5 \times 10^6$ timesteps, respectively. **d.** Relative entropy $D_{KL}(q||p^{design})$ between empirical distribution $q(M)$ and target distribution $p^{design}(M)$ without confinement (blue circles) and under confinement (red triangles). $D_{KL}(q||p^{design})$ also decreases with time following a power law under confinement, where the best-fitting exponent is $\alpha = 1.82 \pm 0.04$,

based on the result of one simulation using data points where $D_{KL}(q||p^{design}) < 0.5$. The scaling exponent is determined by linear regression in the log-log space. **e.** The microscopic evolution during network adaptation. Initially, there are numerous forbidden edges (red lines, left panel). The ratio of forbidden edges is 81.44% at $t = 1$, 72.86% at $t = 41$, 66.83% at $t = 81$ and 60.40% at $t = 121$. During evolution, the forbidden edges disappear and network still exhibits desired bistability (right panel). Here, nodes in the same community are shown in the same colors. Nodes with more connections are larger.

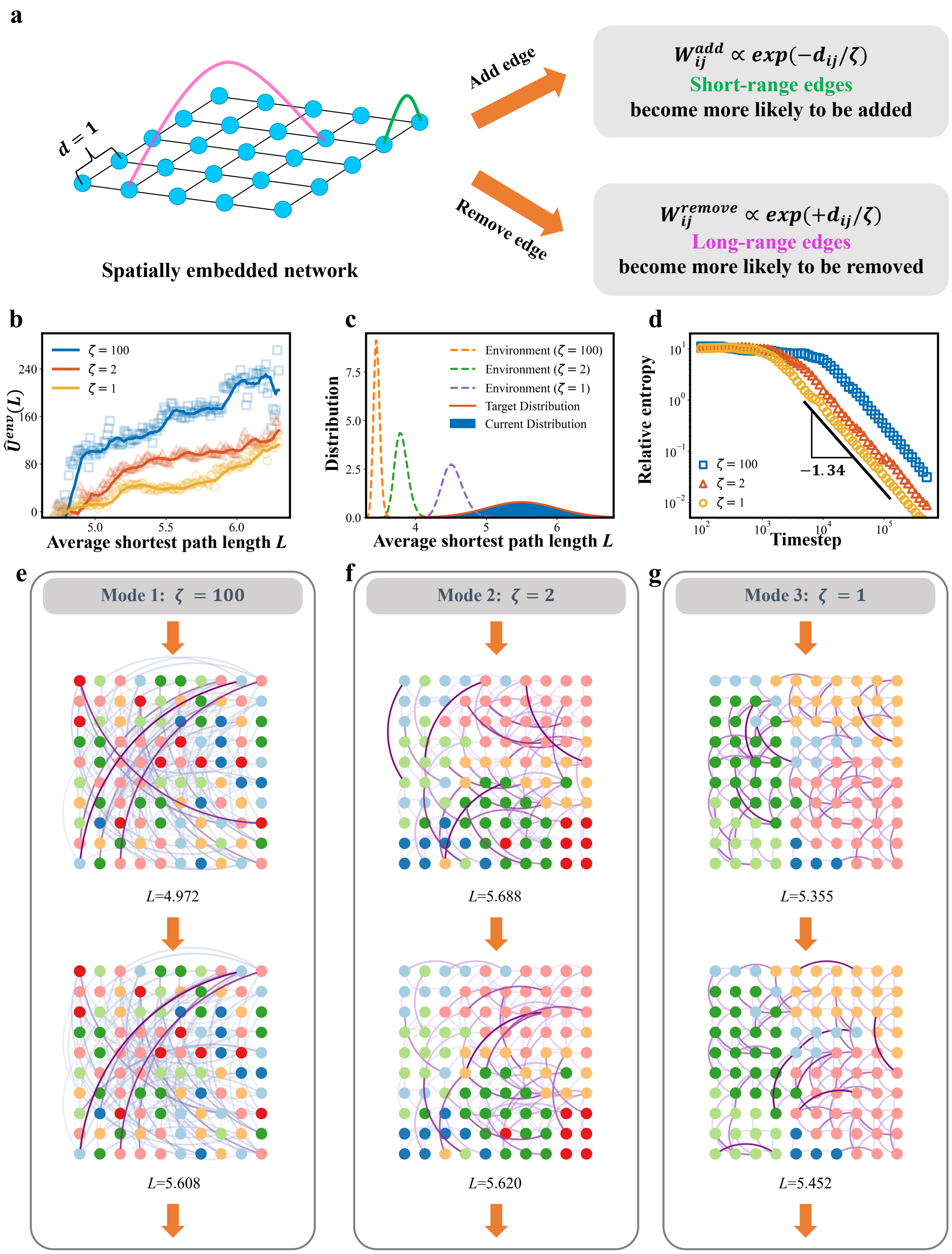

**Fig. 5| Adaptation under geographic constraint. a.** Nodes are embedded in a two-dimensional square lattice, where distance between two neighbors are set to 1. The characteristic length $\zeta$ represent the strength of geographic constraint (see environment setting in **Methods**). **b**. Environment estimation $\widehat{U}^{env}(L)$ obtained in $5 \times 10^5$ timesteps under geographic constraint $\zeta = 100$, 2 and 1. To visualize the trend of $\widehat{U}^{env}(L)$, we smoothed raw data (hollow points) using a moving average method with a window size of 10 to obtain the curve (solid lines). The value of $\widehat{U}^{env}(L)$ at $L =$ 4.805 is adjusted to zero by adding a constant term for comparing $\widehat{U}^{env}(L)$ under

different geographic constraints. **c**. From different environment-dominated distributions (dashed lines), self-adaptive network finally realizes target distribution (solid line). The empirical distributions before and after employing self-adaptation strategy are all obtained by simulating $5 \times 10^5$ timesteps. **d**. Relative entropy $D_{KL}(q||p^{design})$ between empirical distribution and target distribution decreases with time following a power law. The best-fitting exponents are $\alpha = 1.42 \pm 0.01$ for $\zeta = 100$, $\alpha = 1.29 \pm 0.02$ for $\zeta = 2$, and $\alpha = 1.34 \pm 0.005$ for $\zeta = 1$. All the scaling exponents are determined by linear regression in the log-log space, based on the result of one simulation using data points where $D_{KL}(q||p^{design}) < 1$. **e-g**. The microscopic evolution of self-adaptive network under geographic constraint $\zeta = 100, 2, 1$, where nodes in the same community are shown in the same color. Nodes with more connections are larger. Edges with longer distance are darker.

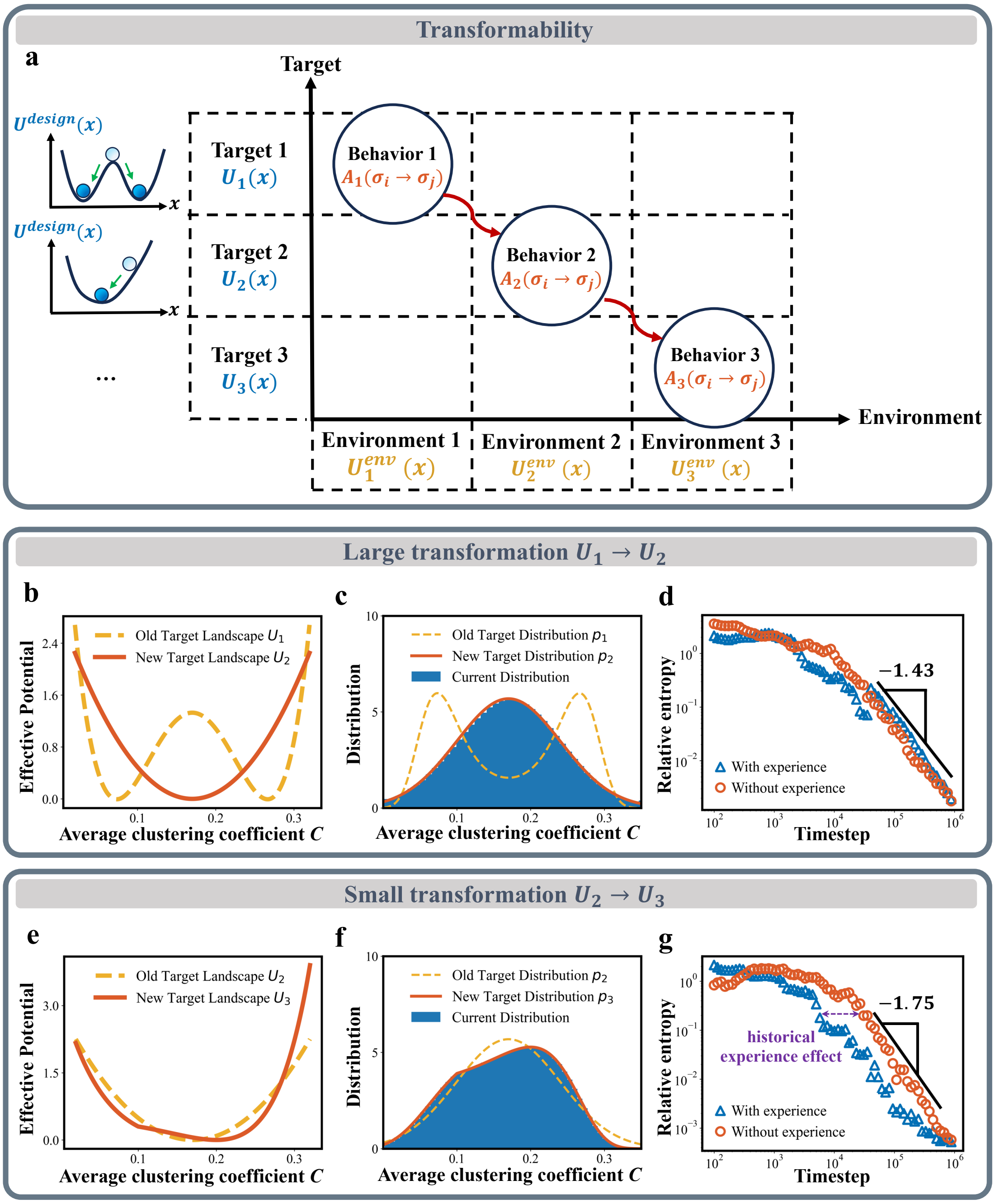

**Fig. 6| Transformability. a.** Illustration of transformability. **b.** Large transformation from bistable landscape $U_1$ to monostable landscape $U_2$. **c.** The distribution of average clustering coefficient $C$ is successfully changed from $p_1(C) \propto \exp(-U_1(C))$ to $p_2(C) \propto \exp(-U_2(C))$ after transformation $U_1 \to U_2$ in $10^6$ timesteps. **d.** The relative entropy $D_{KL}(q||p_2)$ for self-adaptive network with historical experience (blue triangles) and self-adaptive network without experience (red circles) after transformation $U_1 \to U_2$. For self-adaptive network without experience (red circles), the best-fitting exponent $\alpha = 1.43 \pm 0.02$ is determined by linear regression in the log-log space, based on the result of one simulation using data points where $D_{KL}(q||p^{design}) < 0.5$. **e.** Small transformation from landscape $U_2$ to landscape $U_3$. **f.** The distribution of average clustering coefficient $C$ is successfully changed from $p_2(C) \propto \exp(-U_2(C))$ to $p_3 \propto \exp(-U_3(C))$ after transformation $U_2 \to U_3$ in $10^6$ timesteps. **g.** Relative entropy

$D_{KL}(q||p_3)$ for self-adaptive network with historical experience (blue triangles) and self-adaptive network without historical experience (red circles) after transformation $U_2 \rightarrow U_3$. For self-adaptive network without experience (red circles), the best-fitting exponent $\alpha = 1.75 \pm 0.04$ is determined by linear regression in the log-log space gives, based on the result of one simulation using data points where $D_{KL}(q||p^{design}) < 0.5$. The historical experience reduces the time to decrease $D_{KL}(q||p^{target})$ (purple dashed arrow).

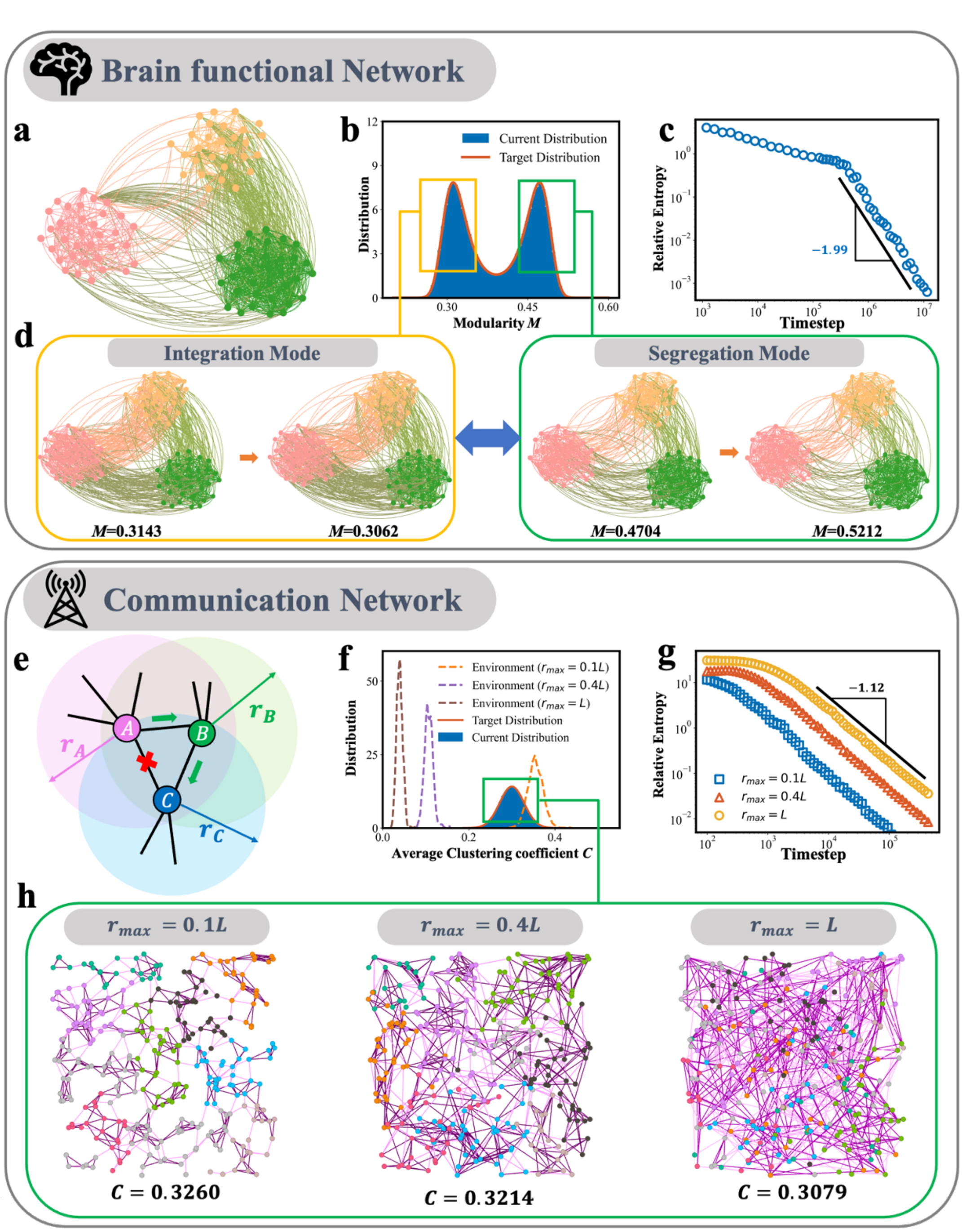

**Fig. 7| Application in real networks. a.** Brain functional network, where nodes in the same community are shown in the same color. See network setting in **Methods**. **b.** The distribution of network modularity $M$ for self-adaptive network after employing self-adaptation strategy in $10^7$ timesteps. **c.** Relative entropy $D_{KL}(q||p^{design})$ between empirical distribution and target distribution decreases with time following a power law, where the best-fitting exponent is $\alpha = 1.99 \pm 0.03$, based on the result of one simulation using data points where $D_{KL}(q||p^{design}) < 0.5$. **d.** The network microscopic evolution after employing self-adaptation strategy. Here, nodes in the same community are shown in the same color. **e.** Communication wireless network, where triangles in network support local fault-tolerance. See network setting in **Methods**. **f.** The distribution of average clustering coefficient $C$ for communication network after

employing self-adaptation strategy in $4.6 \times 10^5$ timesteps. Network is successfully changed from environment-dominated distribution (dashed line) to target distribution $p^{design}(C) \propto \exp(-U^{design}(C))$ (red solid line) under different geographic constraint $r_{max} = 0.1L, 0.4L, L$. The environment-dominated empirical distributions are all obtained by simulating $5 \times 10^4$ timesteps. **g.** Relative entropy $D_{KL}(q||p^{design})$ between empirical distribution and target distribution decreases with time following a power law. The best-fitting exponents are $\alpha = 1.25 \pm 0.01$ for $r_{max} = 0.1L$, $\alpha = 1.16 \pm 0.01$ for $r_{max} = 0.4L$, and $\alpha = 1.12 \pm 0.01$ for $r_{max} = L$. All the scaling exponents are determined by linear regression in the log-log space, based on the result of one simulation using data points where $D_{KL}(q||p^{design}) < 1$. **h.** The network microscopic structure in a snapshot under different geographic constraint $r_{max} = 0.1L, 0.4L, L$. Here, nodes in the same community are shown in the same color. Edges in triangles are drawn darker to highlight them, as they are redundantly supported by others in the same triangles. The ratios of edges in triangles are 72.25%, 73.86%, 65.77% for the snapshot under different geographic constraint $r_{max} = 0.1L, 0.4L, L$, respectively.

**Data Availability**
Network data and simulation code in this study will be released in https://github.com/linearworld/A-simple-intelligent-adaptive-network.

**Acknowledgments**
This work was supported in part by the National Natural Science Foundation of China under Grant 72225012, Grant 72288101 and Grant 71822101, and the Fundamental Research Funds for the Central Universities.

## Methods

### Deriving the suitable acceptance probability

Without any intervention, the evolution of system follows Markov process $T$, where the transition probability from microstate $\sigma_i$ to microstate $\sigma_j$ is denoted as $T(\sigma_i \to \sigma_j)$. And we assume that Markov process $T$ satisfies the detailed balance condition. The detailed balance holds for environment with time-reversal symmetry[59]. Even for systems violating detailed balance, the violation of detailed balance will be reduced after coarse-graining[60]. The detailed balance condition is given by

$$p^{env}(\sigma_i)T(\sigma_i \to \sigma_j) = p^{env}(\sigma_j)T(\sigma_j \to \sigma_i) \tag{8}$$

for each pair of microstates $\sigma_i$ and $\sigma_j$, where $p^{env}(\sigma)$ is the stationary probability of microstate $\sigma$ for the Markov process $T$. The superscript 'env' denotes that system is evolving by environmental driving $T(\sigma_i \to \sigma_j)$. The stationary distribution of Markov process $T$ for macrostate $x$ is

$$p^{env}(x) = \sum_{x(\sigma_i)=x} p^{env}(\sigma_i). \tag{9}$$

Our goal is to change the stationary distribution of macrostate from $p^{env}(x)$ to desired distribution $p^{design}(x)$, where $p^{design}(x) \propto \exp\left(-U^{design}(x)\right)$. To realize this goal, we accept each transition from microstate $\sigma_i$ to microstate $\sigma_j$ with acceptance probability $A(\sigma_i \to \sigma_j)$. Thus system transition probability from microstate $\sigma_i$ to microstate $\sigma_j$ is changed from $T(\sigma_i \to \sigma_j)$ to $T(\sigma_i \to \sigma_j)A(\sigma_i \to \sigma_j)$, leading to the change of stationary distribution. Below we derive the suitable $A(\sigma_i \to \sigma_j)$ for changing system stationary distribution of macrostate to $p^{design}(x)$.

We denote the desired stationary distribution of microstate as $p^{design}(\sigma_i)$, which needs to satisfy

$$p^{design}(x) = \sum_{x(\sigma_i)=x} p^{design}(\sigma_i). \tag{10}$$

To guarantee that distribution $p^{design}(\sigma_i)$ is stationary under transition probability $A(\sigma_i \to \sigma_j)T(\sigma_i \to \sigma_j)$, a sufficient condition is detailed balance condition that

$$A(\sigma_j \to \sigma_i)T(\sigma_j \to \sigma_i)p^{design}(\sigma_j) = A(\sigma_i \to \sigma_j)T(\sigma_i \to \sigma_j)p^{design}(\sigma_i) \tag{11}$$

To satisfy Eq. (11), we choose

$$A(\sigma_i \to \sigma_j) = \min\left(1, \frac{T(\sigma_j \to \sigma_i)p^{design}(\sigma_j)}{T(\sigma_i \to \sigma_j)p^{design}(\sigma_i)}\right) \tag{12}$$

for each transition $\sigma_i \to \sigma_j$. Substituting Eq. (8) into Eq. (12), we get

$$A(\sigma_i \to \sigma_j) = \min\left(1, \frac{p^{design}(\sigma_j)}{p^{env}(\sigma_j)} / \frac{p^{design}(\sigma_i)}{p^{env}(\sigma_i)}\right). \tag{13}$$

Thus we only need to determine the ratio $\frac{p^{design}(\sigma_i)}{p^{env}(\sigma_i)}$ for each microstate $\sigma_i$. The acceptance probability for realizing designed microstate distribution in Eq. (12) has been derived in Markov chain Monte Carlo. Below we extend it to designed macrostate distribution. Even the macrostate distribution $p^{design}(x) \propto \exp(-U^{design}(x))$ is determined, there is a lot of freedom to select the desired microstate distribution $p^{design}(\sigma_i)$, which is only constraint by Eq. (10). We will choose the most convenient $p^{design}(\sigma_i)$ for calculating the ratio $\frac{p^{design}(\sigma_i)}{p^{env}(\sigma_i)}$ as following.

Considering ratio $\frac{p^{design}(\sigma_i)}{p^{env}(\sigma_i)}$ also needs to satisfy the constraint shown in Eq. (10). To get a more convenient form of this constraint, we divide Eq. (10) by Eq. (9) and get

$$\frac{p^{design}(x)}{p^{env}(x)} = \frac{\sum_{x(\sigma_i)=x} p^{design}(\sigma_i)}{\sum_{x(\sigma_i)=x} p^{env}(\sigma_i)}. \tag{14}$$

Thus $p^{design}(\sigma_i)$ only need to satisfy Eq. (14) to satisfy constraint shown in Eq. (10). For convenience of calculating ratio $\frac{p^{design}(\sigma_i)}{p^{env}(\sigma_i)}$ under this constraint, we choose desired microstate distribution $p^{design}(\sigma_i)$ satisfying

$$\frac{p^{design}(\sigma_i)}{p^{env}(\sigma_i)} = \frac{p^{design}(x_i)}{p^{env}(x_i)}, \tag{15}$$

where $x_i = x(\sigma_i)$ is the macrostate of microstate $\sigma_i$. According to Eq. (15), when probability of macrostate $x_i$ needs to be scaled by a factor $\kappa$ ($\frac{p^{design}(x_i)}{p^{env}(x_i)} = \kappa$), the probability of microstates $\sigma_i$ with macrostate $x(\sigma_i) = x_i$ will also be scaled by factor $\kappa$ ($\frac{p^{design}(\sigma_i)}{p^{env}(\sigma_i)} = \kappa = \frac{p^{design}(x_i)}{p^{env}(x_i)}$), which guarantees the satisfaction of Eq. (14).

Substituting Eq. (15) into Eq. (13), we get

$$A(\sigma_i \to \sigma_j) = \min\left(1, \frac{p^{design}(x_j)}{p^{env}(x_j)} / \frac{p^{design}(x_i)}{p^{env}(x_i)}\right) \tag{16}$$

where $x_i = x(\sigma_i)$ and $x_j = x(\sigma_j)$. Notably, the acceptance probability only use the macroscopic information of macrostate $x_i$ and $x_j$. For convenience, we represent probability distribution with free energy landscape $U^{design}(x)$ and $U^{env}(x)$ shown in Eqs. (2-3), then the acceptance probability becomes

$$A(\sigma_i \to \sigma_j) = \min\left(1, \frac{\exp(U^{design}(x_i) - U^{design}(x_j))}{\exp(U^{env}(x_i) - U^{env}(x_j))}\right). \tag{17}$$

With our target $U^{design}(x)$ and environment property $U^{env}(x)$, we can steer system into desired macroscopic landscape $U^{design}(x)$ with only macroscopic information $x_i = x(\sigma_i)$ and $x_j = x(\sigma_j)$.

**Deriving the estimation update method**

As shown in Eq. (4), to realize target landscape with only macroscopic information,

system needs to know the $U^{design}(x)$ and $U^{env}(x)$ for proper decision. $U^{design}(x)$ is system target that is initially known. $U^{env}(x)$ encodes the environment property, which is usually unknown. System needs to use historical data to estimate $U^{env}(x)$. In our strategy, system use its estimation $\widehat{U}^{env}(x)$ to replace $U^{env}(x)$ in Eq. (4). When estimation $\widehat{U}^{env}(x)$ and $U^{env}(x)$ differ only by a constant term ( $\widehat{U}^{env}(x) = U^{env}(x) + const$), the estimation $\widehat{U}^{env}(x)$ will give correct acceptance probability based on Eq. (4) due to the fact that $\widehat{U}^{env}(x_j) - \widehat{U}^{env}(x_i) = U^{env}(x_j) - U^{env}(x_i)$. Algorithms for estimating entropy, including the Wang-Landau method, have been developed in statistical physics while not for realizing various target landscape $U^{design}(x)$. Due to the correspondence between entropy and $U^{env}(x)$, below we extend the Wang-Landau method to derive a simple estimation update method, which enables system to continuously update estimation $\widehat{U}^{env}(x)$ while pursuing target landscape $U^{design}(x)$.

System stores an value of $\widehat{U}^{env}(x_i)$ for each macrostate $x_i$. And the estimation $\widehat{U}^{env}(x)$ will be updated according to observation during system evolution. Inspired by the Wang-Landau method[44,45], when system enters into macrostate $x_i$, we assume that the corresponding estimation $\widehat{U}^{env}(x_i)$ will be updated as

$$\widehat{U}^{env}(x_i) = \widehat{U}^{env}(x_i) - f \times h(x_i), \quad (18)$$

where $f$ is the adaptation rate and $h(x_i)$ is an undetermined function about macrostate $x_i$. And estimation $\widehat{U}^{env}(x)$ on other macrostate $x \neq x_i$ will not change. Next we derive the form of $h(x_i)$ based on stationary condition that correct estimation should not be changed with estimation update method Eq. (18). Estimation $\widehat{U}^{env}(x)$ is correct if and only if estimation $\widehat{U}^{env}(x)$ and $U^{env}(x)$ differ only by a constant term ($\widehat{U}^{env}(x) = U^{env}(x) + const$). After a time interval $\Delta t$, the estimation $\widehat{U}^{env}(x)$ is updated to $\widehat{U}^{env}(x) + \Delta\widehat{U}^{env}(x)$, where $\Delta\widehat{U}^{env}(x)$ is the increment induced by Eq. (18). The increment $\Delta\widehat{U}^{env}(x)$ should be independent of $x$ to keep correct estimation. According to Eq. (18), the increment is

$$\Delta\widehat{U}^{env}(x) = -f \times h(x) \times n_{\Delta t}(x). \quad (19)$$

where $n_{\Delta t}(x)$ is the number of times that system enters macrostate $x$ during the time interval $\Delta t$. When system estimation $\widehat{U}^{env}(x)$ is correct, the probability of entering state $x$ follows the target distribution $p^{design}(x) \propto \exp(-U^{design}(x))$. Thus expected value of increment is

$$< \Delta\widehat{U}^{env}(x) > = -f \times h(x) \times p^{design}(x) \times \Delta t \quad (20)$$

where expected value $<n_{\Delta t}(x)> = p^{design}(x) \times \Delta t$ is used. Considering $\Delta\widehat{U}^{env}(x)$ should be independent of $x$, thus

$$h(x) \propto \frac{1}{p^{design}(x)} \propto \exp(U^{design}(x)). \quad (21)$$

Choosing $h(x) = \exp(U^{design}(x))$, we get the estimation update method

$$\widehat{U}^{env}(x_i) = \widehat{U}^{env}(x_i) - f \times \exp(U^{design}(x_i))). \quad (22)$$

Our method extends the Wang-Landau method. For the constant target landscape $U^{design}(x) = const$, our method degrades into the Wang-Landau method. We prove the convergence of our adaptation rule, consisting of Eq. (4) and Eq. (5), under slow adaptation limit (see Supplementary Section 1). When applying this estimation update method for continuous macrostate $x$, we discretize $x$ into small bins and continuously update its estimation for each bins following Eq. (5). And estimation $\widehat{U}^{env}(x)$ for each macrostate $x$ is determined by linear interpolation between neighboring bins similar to reference[61].

**Environment setting**

As shown in Fig. 2a, the environmental disturbance will drive system from one microstate $\sigma_i$ to another microstate $\sigma_j$ with transition probability $T(\sigma_i \to \sigma_j)$, where $T(\sigma_i \to \sigma_j)$ is unknown for intelligent adaptive network. The environment setting is essentially a procedure to generate a perturbed network $\sigma_j$ from the original network $\sigma_i$. Below we introduce the environment setting used in examples.

Each synthetic network is initialized as a Erdos-Renyi network with $N = 100$ nodes and average degree $< k >= 4$. Environmental disturbance will remove or add edge to network at each timestep. The environmental disturbance setting is constructed as following. At each timestep, environment will first decide whether to add one edge to network or to remove one edge from network. The probability of removing one edge at one timestep is

$$p_{remove} = \frac{1}{1+\exp(-A_m(m-m_0))}, \tag{23a}$$

where $m$ is the number of existing edges in current network. The probability of adding one edge $p_{add} = 1 - p_{remove}$. In the example of synthetic networks, parameters are set to $A_m = 2$ and $m_0 = 200$. In the example of brain functional network, parameters are set to $A_m = 2$ and $m_0 = 990$. When $m < m_0$, environment tends to add edges. When $m > m_0$, environment tends to remove edges. Thus the number of edges in network will fluctuate around $m_0$. In the example of communication network, the probability of removing one edge at one timestep is set to

$$p_{remove} = \exp\left(-B_m \frac{m_{potential}}{m}\right), \tag{23b}$$

where $m_{potential}$ is the number of potentially addable edges under the degree constraint, and $m$ is the number of existing edges in current network. The probability of adding one edge $p_{add} = 1 - p_{remove}$. The parameter is set to $B_m = \frac{\ln(2)}{0.2}$, which guarantees that $p_{remove} = p_{add}$ when $\frac{m_{potential}}{m} = 0.2$.

After deciding whether to add one edge to network or to remove one edge from network, environment selects a specific edge to add or remove. We try different selection methods to test the adaptability of our intelligent adaptive network. In the first example, fourth example and the example of brain functional network, all edges have the same probability of being selected for addition or removal.

In the second example, forbidden edges will not be selected for addition and all existed edges still have the same probability of being selected for removal. After forbidding 80% of all possible edges, the number of possible edges to be added is $20\% \times \frac{N(N-1)}{2}$. For example, we consider a network with $m$ edges. Without confinement, the number of possible microstates is $\binom{\frac{N(N-1)}{2}}{m}$. Under confinement, the number of possible microstates will be reduced from $\binom{\frac{N(N-1)}{2}}{m}$ to $\binom{20\% \times \frac{N(N-1)}{2}}{m}$. The ratio is

$$\frac{\binom{20\% \times \frac{N(N-1)}{2}}{m}}{\binom{\frac{N(N-1)}{2}}{m}} \approx \frac{(20\% \times \frac{N(N-1)}{2})^m}{(\frac{N(N-1)}{2})^m} = (\frac{1}{5})^m. \tag{24}$$

According to Eq. (23a), the number of edges $m$ fluctuates around $m_0 = 200$. Thus the possible microstates of network will be reduced to $5^{-200}$ of original one.

In the third example, the probability of selecting edge (*i,j*) depends on the geographic distance $d_{ij}$ between node *i* and node *j*. When environment selects one edge to add, the probability of selecting edge with length $d_{ij}$ is proportional to $\exp(-d_{ij}/\zeta)$. When environment selects one edge to remove, the probability of selecting edge with length $d_{ij}$ is proportional to $\exp(+d_{ij}/\zeta)$ (Fig. 5a). Here, characteristic length $\zeta$ represents the strength of geographic constraint. The edge with length $d_{ij} \gg \zeta$ is easy to be removed while hard to be added. As $\zeta$ decreases, the geographic constraint will be strengthen.

In all examples, if removing selected edge will break network into disconnected parts, such removal will be directly rejected to ensure network connectivity for calculating average shortest path length. Besides the above environment setting, we test our self-adaptive network in a more generalized environment, where environment may simultaneously add and remove multiple edges (see Supplementary Section 5).

**Target setting**

In the first and second example, the macrostate of interest is chosen as modularity, which is defined as

$$M = \frac{1}{2m} \sum_{i \neq j} (A_{ij} - \frac{k_i k_j}{2m}) \delta(c_i, c_j), \tag{25}$$

where $m$ is the number of existing edges in network. The adjacent matrix $A_{ij} = 1$ if there is one edge between node $i$ and node $j$, otherwise $A_{ij} = 0$. The degree $k_i = \sum_{l \neq i} A_{il}$ and $k_j = \sum_{l \neq j} A_{jl}$ is the number of edges for node $i$ and node $j$, respectively. $\delta(c_i, c_j) = 1$ if node $i$ and node $j$ belong to the same community, otherwise $\delta(c_i, c_j) = 0$. Here, we use Clauset-Newman-Moore heuristic[46] to detect community. The target landscape $U^{design}(M)$ is set to the variation of Landau's free energy function[62]

$$U(M) = a_4 M^4 - a_3 M^3 + a_2 M^2 - a_1 M + a_0, \tag{26}$$

where $a_4 = 5859.375$, $a_3 = 13750$, $a_2 = 11906.25$, $a_1 = 4505.111$ and $a_0 =$

627.442. The target landscape $U(M)$ exhibits two local minima at $M_1 \approx 0.458$ and $M_2 \approx 0.715$. These minima are separated by a local maximum at $M_0 \approx 0.587$.

In the third example, the macrostate of interest is chosen as average shortest path length $L$, which is defined as

$$L = \frac{\sum_{i \neq j} d_{ij}}{N(N-1)}, \tag{27}$$

where $d_{ij}$ is the minimum number of edges that must be traversed to travel from node $i$ to node $j$. The number of nodes is $N = 100$. The target landscape $U^{design}(L)$ of average shortest path length is set to

$$U(L) = 2 \times (L - 5.5)^2, \tag{28}$$

which exhibits only one local minimum at $L = 5.5$.

In the final example, the macrostate of interest is chosen as average clustering coefficient $C$, which is defined as

$$C = \frac{1}{N} \sum_{i=1}^{N} \frac{\sum_{j>k} A_{ij} A_{jk} A_{ki}}{k_i(k_i - 1)/2} \mathbb{1}\{k_i > 1\}, \tag{29}$$

where the indicator function $\mathbb{1}\{k_i > 1\} = 1$ only when the degree $k_i = \sum_{l \neq i} A_{il} > 1$, otherwise $\mathbb{1}\{k_i > 1\} = 0$. The number of nodes is $N = 100$. The sum $\sum_{j>k} A_{ij} A_{jk} A_{ki}$ is the number of triangles including node $i$, and the product $k_i(k_i - 1)/2$ is the numbers of triplet (sets of three nodes with at least two connection) connected by node $i$. The first target landscape is set to the variation of Landau's free energy function

$$U_1(C) = b_4 C^4 - b_3 C^3 + b_2 C^2 - b_1 C + b_0 \tag{30a}$$

where $b_4 = 15432.099, b_3 = 10493.827, b_2 = 2388.889, b_1 = 205.68$ and $b_0 = 5.922$. The target landscape $U_1(C)$ exhibits two local minima at $C \approx 0.074$ and $C \approx 0.266$. These minima are separated by a local maximum at $C \approx 0.17$.

The second target landscape is set to a symmetrical landscape

$$U_2(C) = 100(C - 0.17)^2, \tag{30b}$$

which exhibits only one local minimum at $C = 0.17$.

The third target landscape is set to an asymmetrical landscape

$$U_3(C) = 30 e^{10 \times |C - 0.1|} (C - 0.2)^2, \tag{30c}$$

which exhibits only one local minimum at $C = 0.2$.
Given target landscape $U^{design}(x)$, our self-adaptive network could realize target distribution

$$p^{design}(x) = \frac{\exp(-U^{design}(x))}{\sum_{y \in \Omega_x} \exp(-U^{design}(y))}, \tag{31}$$

where $\Omega_x$ is the operational domain of macrostate $x$.

In the example of the brain functional network, the macrostate of interest is chosen as modularity, where community structure for the modularity calculation is detected based on the original brain functional network using the Clauset-Newman-Moore heuristic. The target landscape is set to

$$U(M) = a_4 M^4 - a_3 M^3 + a_2 M^2 - a_1 M + a_0, \quad (32)$$

where $a_4 = 38400$, $a_3 = 60160$, $a_2 = 34848$, $a_1 = 8840.178$ and $a_0 = 827.591$. The target landscape $U(M)$ exhibits two local minima at $M_1 \approx 0.311$ and $M_2 \approx 0.472$. These minima are separated by a local maximum at $M_0 \approx 0.392$.

In the example of communication network, the macrostate of interest is chosen as average clustering coefficient $C$. The target landscape is set to

$$U(C) = 625(C - 0.3)^2, \quad (33)$$

which exhibits only one local minimum at $C = 0.3$.

**Brain functional network**

The brain functional networks were derived from the HCP database ( https://db.humanconnectome.org/). We used a parcellation provided by Schaefer et al[63] on the fsLR-32k space at resolution 100 regions of interests (ROIs) across both hemispheres. For each ROI, the resting-state time-course was extracted. We calculated Pearson correlation coefficient of the time series between any pair of ROIs. To construct the functional network, we threshold the network to fixed edge density of 20%. Then we remove the nodes out of the largest connected cluster. We ignore edge weights and keep only the topology. This unweighted network consists of 95 nodes and 990 edges. We identify 3 communities in this network with the Clauset-Newman-Moore heuristic shown in Fig. 7a. We use this brain functional network as the initial network configuration in the simulation. The network data will be released in https://github.com/linearworld/A-simple-intelligent-adaptive-network.

**Communication network**

In our wireless communication network, nodes represent sensors, base stations and other equipment, while edges represent the direct communication links between nodes. This wireless communication network consists of 300 nodes randomly deployed within a square area of side length $L$. The wireless communication is constrained by factors such as transmit power, propagation loss and received power[64]. Previous theoretical analyses on communication networks often simplify this constraint by assuming that two nodes can communicate if their distance is below a certain threshold[65], which presumes homogeneous nodes. We extend this model to a more heterogeneous scenario, where nodes have varying transmit power, though receiver parameters remain consistent. Node $i$ can transmit signals to other nodes within a range of $r_i$. Therefore, two nodes, node $i$ and node $j$, can communicate directly if and only if their distance $d_{ij}$ satisfies $d_{ij} \leq \min(r_i, r_j)$. The range $r_i$ is uniformly sampled from the interval $[r_{min}, r_{max}]$ for each node, where $r_{min} = 0.1L$ and $r_{max}$ is set to different values to create environments with different geographical constraint. Here, $r_{max} = 0.1L, 0.4L, L$. Another constraint is on the degree of node due to hardware or protocol factor. The number of direct connections $k_i$ for node $i$ needs to satisfy $k_i \leq k_i^{max}$. Considering the heterogeneity, the upper bound $k_i^{max}$ is randomly sampled from the integers between 4 and 10. The initial communication network topology is constructed by first connecting all edges that satisfy $d_{ij} \leq \min(r_i, r_j)$, and then removing edges until the constraint on degree $k_i \leq k_i^{max}$ is met.